\documentclass[12pt]{iopart}
\usepackage{datetime}

\usepackage{iopams}
\usepackage{epsf}
\usepackage{epsfig}
\usepackage{color}
\usepackage{psfrag}
\usepackage{verbatim}
\usepackage{setstack}
\usepackage{amsopn}
\usepackage{appendix}
\usepackage{graphicx}
\usepackage[small]{caption}
\newcommand{\be}{\begin{eqnarray}}
\newcommand{\ee}{\end{eqnarray}}

\begin{document}

\unitlength = 1mm
\eqnobysec

\title{Coarsening phenomena}
\author{L. F. Cugliandolo}

\address{Sorbonne Universit\'es, 
Universit\'e Pierre et Marie Curie -- Paris VI, 
Laboratoire de Physique Th\'eorique et Hautes Energies,
4 Place Jussieu, 
75252 Paris Cedex 05, France
}

\begin{abstract}
This article gives a short description of pattern formation and coarsening phenomena
and focuses on recent experimental and theoretical advances in these fields. 
It serves as an introduction to phase ordering kinetics and it will appear
in the special issue  `Coarsening dynamics',  Comptes
Rendus de Physique, edited by F. Corberi and P. Politi.
\end{abstract}

\today \hspace{0.5cm} 

\section{Introduction}
\label{sec:intro}

When a parameter in a macroscopic homogeneous system is changed, the homogeneous state 
often becomes unstable and the system evolves into a time-dependent inhomogeneous configuration
with growth of order. The asymptotic selection of a persistent length associated to a given {\it pattern}, or {\it coarsening} with an 
increase in the length scale with time  only arrested by the full homogenisation of the sample, 
are the two major alternatives. 

Problems of this kind are found  in, essentially, all branches of science. In physics, 
phase ordering kinetics occur at very different scales ranging from clustering dynamics in the early universe to the growth of 
macroscopic structures. {\it Coarsening} and {\it pattern formation} are observed in solids as well as in fluids. 
In social sciences, models for, {\it e.g.}, opinion dynamics involve coarsening 
phenomena. In developmental biology,  pattern formation describes the mechanism whereby initially equivalent 
cells in a developing tissue in the embryo assume complex forms and functions. Bacterial colonies grow into 
fancy spatial patterns depending on the surrounding conditions. Many other examples could be used to 
illustrate these phenomena.

Our understanding of coarsening and pattern formation have progressed following 
two main routes. One is theoretical and the main actors have been mathematicians, mostly specialised in 
probability methods, theoretical physicists and theoretical chemists. The other one is experimental, with important contributions
made by the applied physics community. Indeed, in the latter field of research the 
interest in phase ordering kinetics lies on the fact that material properties  usually depend upon the morphology of the phase 
separating sample. In metallurgy, a rather old domain of material science, the motivation for blending 
immiscible components is to create materials with interpenetrating domains
that can have better properties than their components. In more recent material engineering, 
creating micro-patterns and nano-patterns of polymeric materials  is of great interest for 
microdevice fabrication. 

Phase ordering kinetics occur in systems in relaxation
but it also exists under external drive.  The microscopic dynamics could follow physical rules and respect 
detailed balance allowing the system to reach thermal  equilibrium asymptotically, or 
they could go beyond this limitation, in models for social sciences, economy or other non-physical 
branches of science.

The dynamics across (standard) phase transitions are rather well understood qualitatively 
although a full quantitative description of
specific cases is hard to develop. Different mechanisms for pattern formation and 
coarsening exist and have been studied in great detail;
let us mention a few of them here.

We  focus first on the freely relaxing case with dynamics allowing for equilibration.
In the late stages of coarsening, the configurations within domains of, say,  two equilibrium
phases, have already reached equilibrium and the excess free-energy of the system, that is now 
localised at the interface between the two phases, decreases with time. 

In systems with {\it non-conserved scalar order parameter} as realised, for instance, in magnetic systems,
our analytical understanding of phase ordering kinetics is based on the dynamics of these interfaces
and the driving force behind coarsening is just surface tension. This kind of dynamics is often referred
to as {\it curvature driven}.

The simplest example of dynamics with {\it conserved order parameter} is 
Ostwald ripening, the phenomena whereby smaller particles in solution dissolve and deposit on larger 
particles in order to reach a more thermodynamically stable state in which the surface to volume ratio is minimised. 
As the larger particles grow, 
the area around them is depleted of smaller particles. This phenomenon was originally observed in solid solutions
but is also common in emulsions, 
such as the oil-in-water one, and many others. 

In phase separating mixtures, another realisation of conserved order parameter dynamics, 
two possible mechanisms for the initial domain formation exist:
\begin{itemize}
\item[-] 
If the quench ends at a point outside the spinodal curve, the system is
stable against small fluctuations. Rare large fluctuations
of thermally activated origin
create Ôcritical dropletsÕ (nucleation) which subsequently coarsen (growth).
\item[-]
If the quench ends at a point inside the spinodal curve, the system is
unstable against small fluctuations, leading to phase separation by unstable
growth.
\end{itemize}
In the late stages of growth, both mechanisms give rise to qualitatively similar
coarsening motifs. Depending upon the relative density of the two components
one can find isolated droplets of the minority phase immersed in the majority one, 
with their further evolution following Ostwald ripening, 
or a bi-continuous structure with domains of the two kinds percolating across the sample.
In these systems the curvature sets up a gradient in the chemical potential that 
causes molecules to diffuse from regions of high positive curvature to regions of low or negative curvature. 
This is an evaporation-condensation 
mechanism in which the minority-phase domains grow by molecular diffusion through the majority phase.
Segregation in binary alloys is driven by diffusion while  the asymptotic behaviour of segregation in 
binary fluids  is drastically modified by flow fields.

The excess free-energy in a quenched system with scalar order parameter
is stored in sharp domain walls. In cases in which the order parameter is a vector, other kinds 
of topological effects exist. During the ordering process the system has to eliminate some of them to 
reach configurations typical of equilibrium over larger and larger scales. This occurs, for 
instance, in the $2d$ xy model quenched below its Kosterlitz-Thouless temperature,
with the progressive annihilation of vortex-antivortex pairs in the course of time.

The  mechanisms mentioned above are not the only ones leading to coarsening. In non-physical 
problems in which the microscopic dynamics do not necessarily respect detailed balance, 
the evolution may lead to absorbing states {\it via} a coarsening process in which the interfaces are 
of totally different nature and do not have surface tension. This is the case of, {\it e.g.},  the voter model.

I do not intend  to write here a complete overview on the theoretical description of 
pattern formation and coarsening phenomena as several excellent review articles, 
{\it e.g.}~\cite{Cross-Hohenberg,Hohenberg-Halperin,Bray94,Bray03,Puri09-article,Cates12}, and 
textbooks~\cite{Onuki02,Puri09,KrapivskyRednerBenNaim10,HenkelPleimling10} treating these topics already exist, 
and many technical aspects will be covered in the following articles. I will, instead, start by giving a very short 
survey of the main features of phase ordering kinetics addressing the reader to already published 
references and the relevant chapters in this volume for further details.  I will then describe some open 
questions and recent methods that are currently 
being explored and that I personally find interesting and promising.

\section{Experiments}

Traditionally, coarsening systems are studied in the lab with scattering methods. These
give access to the time-dependent structure factor, $S(k,t)$, and the dynamic scaling 
assumption was formulated to describe the time and wave-vector dependence of this 
observable. More subtle features, such as the Porod tails characterising the long wave-vector 
dependence of $S$, were also uncovered from this kind of experimental measurement~\cite{Bray94,Puri09-article}.
In spite of the success of Fourier-space methods, real-space techniques are better suited to give direct access to 
the microscopic mechanisms for coarsening and, therefore, a more detailed understanding of these phenomena. Indeed, a variety of 
visualisation methods that are currently being developed should allow one to observe 
different aspects of coarsening systems in much more detail that what has been done so far.

Direct imaging of the domain structure is easier to achieve in two dimensional systems. 
For instance, temperature controlled polarising microscopy was used in~\cite{Dierking,Sicilia-etal08}
to study the chiral domain structure in electric field driven deracemization of an achiral liquid crystal.
Optical microscopy is commonly used to study domain growth in polymer thin films~\cite{HaasTorkelson97}.
Magnetic force microscopy, photoemission electron microscopy and Lorentz transmission electron microscopy
are used to visualize the magnetic moment configurations in 
artificial magnetic materials, such as the $2d$ spin-ice samples~\cite{artificial-spinice,Heyderman13} that undergo two kinds of order-disorder phase transitions
across which stripe magnetic ordering develops~\cite{Budrikis11,Levis12,Wysin-etal11,Levis13}.

New experimental techniques make now possible the direct visualisation of the domain 
structure of {\it three-dimensional} coarsening systems as well. In earlier studies, the 
domain structure was usually observed {\it post mortem} and only on two-dimensional slices of the 
samples with optic or electronic microscopy. 
Nowadays, it became possible to observe the full $3d$ micro structure {\it in situ},  in the course of evolution.

Three dimensional images give, in principle, access to a complete topological characterisation of the 
interfaces via the calculation of quantities such as the Euler characteristics and the local mean and Gaussian curvatures. 
On top of these very detailed analyses, one can also extract the evolution 
of the morphological domain structure on different planes across the samples and investigate 
up to which extend the third dimension has an effect on what occurs in strictly two dimensions. 
These methods, and the study of the structures in real space, are becoming more and more popular. We here
mention just a few applications to different kinds of domain growth systems.  

In the context of  soft-matter systems, laser scanning confocal microscopy 
was applied to  phase separating binary liquids~\cite{WhiteWiltzius95} and polymer blends~\cite{Jinnai95,Jinnai97,Jinnai99}.
For example, from images at a very late stage of phase separating bicontinuous polymer blend
made of 50 \% polybutadiene and 50\% polyisoprene, 
Jinnai {\it et al.}~\cite{Jinnai04} observed saddle-shaped surfaces with
the statistical averages, $K_M \simeq 0$ for the mean curvature, and 
$K_G\simeq -6.2 \ 10^{-2} \mu$m$^2$ for the Gaussian curvature. Therefore, the interfaces
resemble, on average, minimal surfaces (where $K_M=0$ at each surface point),
although considerable deviations were also reported.

X-ray tomography, a way to observe slices of the sample in a progressive and non-destructive manner,  
was used to quantify phase separating glass-forming liquid binary mixtures~\cite{Bouttes}. 
Among many other interesting features, this study showed that, unexpectedly, this system evolves in a diffusive hydrodynamic regime.

In the realm of magnetic systems the method presented in~\cite{Manke-etal10} looks very promising.

\section{Models}

Phase ordering kinetics are modelled with kinetic Ising or Potts models~\cite{KrapivskyRednerBenNaim10}, 
lattice Boltzmann simulations~\cite{Cates12,GonnellaYeomans09}, 
and stochastic equations on coarse-grained fields~\cite{Bray94,Puri09-article}. These approaches give access to different 
aspects of the processes, and which one is easier to deal with analytically depends upon the 
issue to explore.

Interacting two-state systems were originally introduced to model magnetic phase transitions in easy axis magnets. 
The two-state variables, Ising spins, can be suitably transformed to occupation numbers and model  with them, {\it e.g.}, binary mixtures, 
they can be generalized to take many values and describe with them, {\it {\it e.g.}}, soap froths and, 
still, the updates between different states can be chosen in very different ways to capture various kinds of microscopic 
dynamics.  

The conservation laws, if any,  should be respected by the microscopic updates. 
The two main universality classes one encounters are: 
\begin{itemize}
\item[-]
Non-conserved order parameter dynamics, with its most prominent 
example being the ordering dynamics of a ferromagnet with scalar of vectorial order parameter. 
\item[-]
Conserved order parameter dynamics, its typical example being 
the kinetics of phase separation of a binary mixture.
\end{itemize}

The master equation is the basic analytic tool to deal with the stochastic dynamics of discrete variables.
This route was opened by R. Glauber in his analysis of the stochastic dynamics of the 
Ising chain and it has been followed by numerous authors in varied 
contexts~\cite{KrapivskyRednerBenNaim10,driven-diffusive,Stinchcombe01,Schutz01}. 
However, in general, the master equations lead to a hierarchy of 
coupled equations for the correlation functions that cannot be disentangled nor solved analytically.
Numerical simulations using Monte Carlo 
methods are, instead, easy to implement and provide us with valuable information.

The other popular way to describe coarsening phenomena is phenomenological and it has 
proven to be very successful and quite complete. It is based on the identification of a coarse-grained 
order parameter, the definition of a deterministic force (that may or may not derive from a thermodynamic Landau free-energy density functional),  
and the proposal that the order parameter evolution is governed by a Langevin equation that respects the conservation laws. 
This leads to the time-dependent Ginzburg-Landau equation for non-conserved
order parameter dynamics, the Cahn-Hilliard equation for the conserved order parameter
case, and other stochastic differential equations.

This approach allows one to reach many interesting conclusions. For instance, 
with the field theory one proves that there is no pinching off of protrusions in the 
$2d$ continuous curvature driven dynamics at zero temperature. Instead, an interface 
in $3d$ that has both convex and 
concave parts can undergo fission if the concave portion is thin enough.
However, a full solution to these problems is hard to develop as they are set into the form of a non-linear
functional Langevin equation with no small parameter.

The microscopic discrete models as well as the field theoretical ones can be extended to include 
energy injection via non-conservative forces, or special microscopic dynamic rules.


\section{Dynamic scaling}

{\it Domain growth}~\cite{Bray94,Puri09-article} in systems with scalar order parameter is characterised by a 
patchwork of large domains the interior of which is basically thermalised in one of the 
equilibrium phases while their boundaries slowly move and tend to become smoother due to their elastic energy. 
The patterned structure is not quiescent, ordered regions grow on average with a linear length 
$R(t)$,  but the time needed to fully order the sample diverges with the system size.

The {\it dynamic scaling hypothesis} states that 
at late times and in the scaling limit $r\gg \xi$,  the system is characterized by a single length-scale
$R(t)$ such that the domain structure looks similar at different times if one rescales lengths by $R(t)$.
In practice, this means that all
 time and space dependencies in correlation functions appear as ratios between distances and $R(t)$. 
The length scale is not sensitive to microscopic details but it is to the dimension of the order parameter, 
the conservation laws and the presence of quenched randomness. 
  In clean systems the characteristic length grows algebraically in time, 
 $R(t) \simeq t^{1/z}$, with $z$ a dynamic exponent that defines the dynamic universality class. 
 In scalar systems with non-conserved order parameter dynamics $z=2$.
 In scalar systems with conserved order parameter dynamics 
 bulk diffusion dominates on long length scales and $R(t) \simeq t^{1/3}$ while surface diffusion
can be important when the scales are
small and $R(t) \simeq t^{1/4}$. Surface diffusion, negligible as $t\to\infty$, often dominates
in the experimental range of interest. 

{\it Hydrodynamic flows} make the dynamics of liquids more complicated~\cite{Onuki02,GonnellaYeomans09}.
Indeed, in a phase separating system the mean local curvature of the interfaces induces a pressure difference that 
produces flow. Therefore, in a binary mixture, fluid flow also contributes to the
transport of the order parameter. When diffusion dominates
domain growth, the growth law reduces to the one of a binary
alloy, $R(t) \simeq t^{1/3}$. In the low Reynolds number limit, a simple argument whereby the friction and Laplace 
pressure terms are asked to be of the same order yields $R(t) \simeq t$ (viscous hydrodynamic regime).
For later times and/or large Reynolds numbers inertial effects on the flow velocity can no longer be neglected 
and  a different order of magnitude argument leads to $R(t) \simeq t^{2/3}$
(inertial hydrodynamic regime). The cross-over between these growth-laws 
was checked with lattice Boltzmann simulations~\cite{GonnellaYeomans09}. 
A discussion of the subtle cross-over between these regimes can be found in~\cite{Bray03,Cates12}.
Coarsening in fluids will be further discussed in the chapter by S. Das, S. Roy and J. Midya.

Hydrodynamic effects of this kind are especially important in polymer systems.
In symmetric polymer blends, where constituent polymers have
nearly identical molecular weights and viscoelastic properties, the hydrodynamic interaction
eventually governs the late-stage phase separation in the same manner as in usual binary
fluid mixtures. In asymmetric polymer blends, however, viscoelastic effects unique to
polymers can drastically influence phase separation, see~\cite{Onuki02,Tanaka12} for more 
details.

In systems with {\it competing interactions} the growth of $R(t)$ can be much slower and logarithmic 
growth has been obtained in some of these cases. These models have been used to mimic 
the slow down in glassy systems~\cite{LesHouches,Tarjus}.

{\it Weak quenched disorder} slows down the dynamics of macroscopic systems with respect to their  clean limit. 
Take, for instance, the random field Ising model.
For probabilistic reasons, the fields can be very strong and positive in some region of the sample and favour positive magnetisation,
and very strong and negative in a neighbouring region and favour negative magnetisation. It will then be very hard to 
displace the phase boundary and let one of the two states conquer the local part. Indeed, in $2d$ the random fields 
destroy the finite $T$ transition while in $3d$ they just deplete the ordered phase.
The probability of finding such rare regions can be quantified 
and the relaxation of a perturbation away from 
equilibrium becomes slower than exponential, the dynamic counterpart of the Griffiths essential 
singularities  of the free-energy. The assumption of a power law dependence of 
free-energy barriers with size combined with an Arrhenius argument suggests $R(t) \simeq \ln^{1/\psi} t$
for the growing length.
  
{\it Strong quenched disorder} renders the dynamics still more complex. 
Spin-glasses are the archetype of such systems.
After a quench from high to  low temperatures they not only show very slow out of equilibrium
relaxation, never reaching thermal equilibrium, but  
they also display very intriguing memory effects under complicated paths in parameter space~\cite{Vincent}.
 Although it is not clear whether the dynamics occur {\it {\it via}} the growth of domains, scaling of dynamic correlation functions 
 describe numerical data quite precisely and, somehow surprisingly, with a power law $R(t)\simeq t^{1/z}$.
 This fact is not compatible with mean-field predictions of a much complex time-dependence
 that could perhaps only establish at much longer time scales. The power-law growth of $R(t)$ is not compatible either with 
the droplet picture predictions. In this model static order is assumed to grow as in standard coarsening systems
with two equilibrium states related by symmetry.
The evolution at low enough temperatures should then be dominated by thermal activation with the typical free-energy 
barrier to nucleate a droplet assumed to scale as $L^\psi$ with $\psi$ a
non-trivial exponent. Dynamical observables should then follow scaling laws in terms of a growing length $R(t) \simeq \ln^{1/\psi} t$.
Before drawing conclusions one must keep in mind that the analysis of experimental and  numerical data is  difficult
given the limited range of time scales available in both cases. In~\cite{Corberi-Cugliandolo-Yoshino} an efficient strategy for data analysis 
with the goal of finding the best $R(t)$ for dynamic scaling is discussed and might be of help in the future analysis of this question.

The chapter by F. Corberi presents a discussion of quenched randomness effects on coarsening systems.

We have already stressed that the dynamics of coarsening systems with scalar order parameter is very much determined by the 
dynamics of the interfaces between domains. Coarsening is therefore closely related to the 
dynamics of elastic manifolds (with or without quenched disorder)~\cite{elastic-lines,Giamarchi,Nattermann,Giam-Kolton,Iguain}. The dynamics of {\it directed manifolds} is
relevant, for instance, to the understanding of crystal or other surface growth and is also a problem of great interest. It 
will be discussed in the chapter by C. Misbah and P. Politi.

In systems with vector order parameter, spin-waves are typically accompanied by topological defects and these 
may have an influence on the growing length~\cite{Bray94,Puri09-article}. The latter is identified from the scaling of the correlation functions.
In the $2d$ xy model with non-conserved order parameter
$R(t) \simeq (t/\ln t)^{1/2}$ and the logarithmic correction is due the vortices while in $3d$ the simple $R(t) \simeq t^{1/2}$
is recovered.

\section{Exact solutions and approximation}

A few completely or partially solvable models for coarsening exist but they are confined to one dimension, as the Glauber Ising 
chain or the one-dimensional Ginzburg-Landau equation,
 or are mean-field approximations that can be 
very simple or quite refined as the large $N$ limit of a vector field stochastic dynamics, 
see {\it e.g.}~\cite{KrapivskyRednerBenNaim10,HenkelPleimling10}.
The analysis of the effect of quenched disorder in the 
form of random fields can be done in $1d$ thanks to the strong disorder renormalisation group method~\cite{LeDoussal}.
In the more realistic finite space and internal dimension problem one cannot find analytical
solutions to the discrete or continuous variable models.

Several approximation schemes, including self-consistent approximations to the perturbation series and 
auxiliary field methods, have been developed over the years to characterize the correlation functions. 
These methods, notably the ones due to 
K. Kawasaki and collaborators and G. Mazenko and collaborators, are well explained in the literature, 
see {\it e.g.}~\cite{Bray94,Puri09-article,Onuki02}. They are quite successful for non-conserved order
parameter dynamics as they predict scaling functions for the space-time correlations 
that are very close to the ones obtained with 
numerical or experimental methods. They are, though, less precise for scalar conserved order parameter~\cite{Onuki02}. The problem of 
finding a good analytic approach to this phenomenon remains, thus, open.

\section{Morphologies}

An ensemble of works focused recently on the geometry of the time-dependent mosaic domain structure 
in Ising models and the eventual metastable states reached at zero temperature. We here 
summarise the main questions posed and the results obtained. I give references to the original articles 
as these results have not been collected nor explained in reviews yet.

\paragraph{Influence of the initial state.}

Two kinds of initial conditions should be distinguished and have a different
subsequent low-temperature dynamics. Initial conditions drawn from equilibrium
at super-critical temperature ($T_0>T_c$) have finite correlation length while this diverges in the 
initial configurations that are typical of the critical point ($T_0=T_c$). The evolution 
of these states are rather different as quantified by the study of the correlation functions~\cite{BrayHumayunNewman},
the time-dependent pattern of domains~\cite{ArBrCuSi07,SiArBrCu07,SiSaArBrCu09} and the probability of getting stuck in 
metastable states at zero temperature~\cite{BaKrRe09,OlKrRe11a,OlKrRe11b,OlKrRe12,BlPi13}.

\paragraph{Percolation in the $2d$ kinetic Ising model.}

It was shown in~\cite{ArBrCuSi07,SiArBrCu07} that, after a very short time span (a few Monte
Carlo steps for the simulated cell),  the morphological and statistical properties of the large structures 
in a $2d$ Ising model quenched from high to low temperature (areas of domains, lengths
of interfaces, etc.) look like the ones of site percolation at its threshold.
As the occupation probability
for up and down spins in a high-$T$ equilibrium configuration is smaller than the one at critical percolation, this
fact suggests that the system must have reached critical
percolation at some point. A careful study of this approach demonstrated that
the time needed to reach critical percolation scales as $t_p(L) \simeq L^{\alpha_p}$ with 
$\alpha_p$ an exponent that is smaller than one, depends upon the lattice structure and, 
presumably, the microscopic dynamics~\cite{BlCoCuPi14}. Corrections to scaling due to this 
time-scale, that can be transformed into a length-scale, have been discussed in this work as well
and should be important for the description of experimental data.

\paragraph{Zero-temperature metastable states.}

In $d=1$ both the kinetic Ising chain with Glauber dynamics and the time-dependent Ginzburg-Landau
equation  with non-conserved order parameter 
ultimately  reach complete alignment in a time-scale that grows algebraically with the 
system size in the former case and exponentially in the system size in the latter. However, 
in the absence of thermal fluctuations, the $d\geq 2$ ferromagnetic coarsening may freezes into metastable states. 

In $d= 2$ the discrete model reaches the ground state or a stripe state with probabilities that depend upon the initial condition
($T_0>T_c$ or $T_0=T_c$) and  are equal, with high numerical precision, to the ones of having a spanning cluster at
critical percolation in the former case~\cite{BaKrRe09,OlKrRe12} and different kinds of critical spanning clusters in the latter
case~\cite{BlPi13}. 

In a time-scale scaling as $L^2$ the continuous model may also reach a stripe state but this one will eventually disappear due to the 
domain-wall interaction to let the system condense into the ground state. This second process takes place in a much longer time-scale, 
that diverges exponentially with the system size and lies well beyond experimentally relevant time-scales
for typical sample sizes.

In $d=3$ the discrete model never reaches the ground state and several aspects of the metastable states have been 
discussed in~\cite{OlKrRe11a,OlKrRe11b}. In the continuous case, the Allen-Cahn equation implies that the 
long-lived metastable states are minimal 
surfaces (with vanishing local mean-curvature) compatible with the boundary conditions. 

For the Kawasaki spin-exchange  dynamics, 
because of the conservation of the magnetization, a system quenched to $T=0$ always gets stuck in a metastable state.

\paragraph{Morphology of the time-dependent mosaic structure.}

The distributions of domain lengths in Glauber Ising and Potts chains have been derived 
analytically~\cite{DerridaZeitak96,KrapivskyBenNaim97,Ouchi-etal07,LeDoussal}. The methods 
used in these papers are not adapted to be applied in higher dimensions. Sill, 
many statistical and geometric properties of the hull-enclosed areas and domains in two-dimensional 
coarsening of Ising and Potts models with non-conserved and conserved order parameter dynamics
have been elucidated~\cite{ArBrCuSi07,SiArBrCu07,SiSaArBrCu09,LoArCuSi10,LoArCu12}.
The results for hull-enclosed areas in the non-conserved order scalar parameter universality
class are, quite surprisingly, exact~\cite{ArBrCuSi07,SiArBrCu07}.
Similar studies of the ferromagnetic Ising universality class on two dimensional slices of the $3d$ 
lattice are on their way~\cite{Arenzon-etal15}. Analysis of the full $3d$ structure obtained numerically 
appeared in~\cite{Holyst01,Holyst02} for non-conserved order parameter dynamics and in~\cite{ArakiTanaka00} 
for viscoelastic phase separation.

\section{Cooling rate dependencies}

In condensed-matter systems most theoretical studies of phase ordering
kinetics focus on the dynamics after infinitely rapid quenches as finite quench timescales
are expected to alter the dynamics only at short times and eventually become irrelevant. Cooling-rate
dependencies in defect dynamics are, instead, central in cosmological theories as their time-scales 
are very different from the ones of the laboratory. Although the scenario whereby topological
defects would act as seeds for matter clustering seems to be excluded by observation, the interest in
predicting their density remains due to other possible effects of these objects. In the late 80s Zurek
assumed that defect dynamics beyond the transition are negligible and used critical scaling to derive a
quantitative prediction for the density of defects; he also proposed to check these ideas in condensed-matter
systems~\cite{Zurek85,Zurek96}. A large experimental and numerical activity followed with variable results
summarized in~\cite{Kibble07}. In two recent papers we revisited the Kibble-Zurek mechanism in infinitely
slow annealing across classical thermal second-order~\cite{Biroli10} and Kosterlitz-Thouless~\cite{Jelic11} phase transitions 
and we elucidated the time and cooling-rate dependencies of the density of topological defects
domain walls or vortices) left over in the low-$T$ phases. Our results proved that the dynamics in the 
low-$T$ phase contribute significantly to the reduction of the density of topological defects in quenches with
dissipative dynamics.  Some other studies along these lines appeared recently in, {\it e.g.}~\cite{Krapivsky10,Liu-etal14}.

\section{Linear response}

The time-dependent scaling properties of linear response functions at
criticality~\cite{Calabrese-Gambassi} and in the ordering phase~\cite{Corberi-etal07}
have been investigated by many authors. The motivation for these studies was
to compare their scaling behaviour to the one of the associated correlation functions 
and to analyse how the equilibrium fluctuation-dissipation relation is modified 
out of equilibrium~\cite{Cugliandolo-Teff}. The ultimate purpose being the comparison to 
the fluctuation-dissipation relations in glassy systems with no obvious coarsening process.
I will not pursue the description of the
scaling forms found as these have been reported in detail in the three review
articles mentioned above.

\section{External drive}

The influence of an external drive such as an imposed shear flow, turbulence, gravity or temperature gradients
have been the object of intensive study due to their relevance to applications. In spite of 
these efforts, basic questions about their effect on coarsening remain still open.

In {\it binary fluids under an external shear flow} two interesting open questions
are: which is the late-time behaviour and whether the coarsening continues
indefinitely or is ultimately arrested. This problem has been treated in some simplified cases.
For example, it was shown that flow imposed by weak shear on two opposite surfaces of a two-dimensional 
binary fluid makes coarsening in the direction of flow more rapid. 
Instead, the behaviour of a binary fluid under strong shear is still an open question, and whether growth 
saturates in the transverse direction or not is still a matter of debate~\cite{Bray03,Onuki02}.

Another externally driven system where coarsening effects appear are {\it vibrated granular
matter}. In these systems a variation of the input energy can 
change the internal ordering of the material and a 
transition from the disordered to the ordered phase may be accompanied by 
the growth  of the size of ordered domains.
Coarsening in granular matter will be discussed in the chapter by 
A. Baldassarri, A.  Puglisi and A. Sarracino.

The constituents of {\it active matter}~\cite{active-matter} absorb energy 
from their environment or internal fuel tanks and use it to carry out motion. Energy is partially transformed into 
mechanical work and partially dissipated to the environment in the 
form of heat. The effect of the motors can be dictated by the 
state of the particle and/or its immediate 
neighbourhood and it is not necessarily fixed by an external field.
The units interact directly or through disturbances propagated in the medium. 
Active matter exists at very different scales including bacterial suspensions
and swarms of different animals. 

Active matter displays out of equilibrium phase transitions that may be absent in their passive counterparts. 
Much theoretical effort has been recently devoted to 
the description of different aspects, such as the self-organisation of living 
microorganisms, the identification and analysis of states with spatio-temporal structure, 
and the study of the rheological properties of active particle suspensions. 

A particularly interesting feature of active matter, in the context of
this special issue, is the spatial phase separation into
an aggregated phase and gas-like regions for sufficiently large
packing fractions in the complete absence of attractive interactions.
The current research in phase separation in active fluids 
will be developed in the chapter by A. Tiribocchi, G. Gonnella and D. Marenduzzo.

{\it White multiplicative noise} can induce highly nontrivial effects in one variable~\cite{Horsthemke84} as well as 
extended systems. On the one hand, convective patterns  are predicted by the Swift-Hohenberg equation near 
deterministic points where no pattern would exist without the external multiplicative white noise~\cite{GarciaOjalvo-etal93}. 
On the other hand, 
phase transitions in problems in which the deterministic part does not exhibit any symmetry breaking
have also been exhibited~\cite{VandenBroeck-etal94,VandenBroeck-etal97}. Indeed, 
a short time instability is generated owing to the noise, and the generated nontrivial
state is afterwards rendered stable by the spatial coupling. The transition shows a  divergence of the correlation length, 
critical slowing down, and scaling properties, similarly to what occurs at a conventional 
second order equilibrium phase transition. The dynamics display coarsening effects~\cite{VandenBroeck-etal97}.

Godr\`eche and collaborators (see, {\it e.g.}~\cite{Godreche} and refs. therein) 
have recently focused on the role played by
{\it spatial asymmetry} in the dynamics of phase ordering systems. In the context of
Glauber dynamics, asymmetry means that the 
flipping spin is more influenced by some of its neighbours than by other ones.
Such an asymmetric dynamics are therefore irreversible, because
the principle of action and reaction is violated and detailed balance no longer holds.
However, for some choices of the updating rules, the dynamics still take the system to the Gibbs-Boltzmann
distribution function asymptotically and a coarsening process can establish below some critical value of the 
control parameter.

Coarsening arises also in the approach to condensation in driven diffusive systems. This phenomenon occurs 
in, {\it e.g.}, zero-range processes, models in which equivalent particles
hop from sites to sites on a lattice, with prescribed rates which only depend on the 
occupation of the departure site. Urn models, models of mass transport, and other driven dissipative models also 
show this phenomenology~\cite{Godreche-urn}. The condensate
manifests itself by the macroscopic occupation of a single site of a thermodynamically
large system by a finite fraction of the whole mass in the first models mentioned, while for driven
diffusive systems, the condensate manifests itself as a domain of macroscopic size.
The coarsening process leading to condensation will be described in the chapter by C. Godr\`eche.

\section{Pattern formation}

\textcolor{black}{\it Pattern formation}
is the spontaneous formation of macroscopic spatial structures in open systems constantly  driven 
far from equilibrium. Patterns can be  stationary in time and periodic in space, 
periodic in time and homogenous in space or periodic in both space and time.
The key words `dissipative structures' or 
`self-organisation' are also attached
to this phenomenon that was initially studied  
in fluid dynamics and chemical reactions but also appears in 
solid-state physics, soft condensed matter and nonlinear optics,
and is now central to biology with, for instance,
morphogenesis and the dynamics of active matter playing a predominant role. 
The traditional example is Rayleigh-B\'enard convection: a fluid placed between two infinite 
horizontal plates
that are perfect heat conductors at different temperature. At a threshold value of the 
temperature difference the uniform state with a linear temperature profile
becomes unstable towards a state with convective flow. In the context of chemistry the 
paradigm are systems with  competition between 
temporal growth rates and diffusivity of the different species.

Systems that form patterns are usually described by over-damped dissipative 
deterministic partial differential equations. The exact form of the 
equations depends on the problem at hand, ranging from Navier-Stokes equations 
for fluid dynamics to reaction-diffusion equations for chemical systems.
A well-studied case is the non-linear Schr\"odinger equation.

The spatio-temporal structures are found from the growth and saturation 
of modes that are unstable when a control parameter is increased beyond
threshold. A parallel between the kind of bifurcation of fixed-point solutions and the 
order of a phase transition can be established.
Concretely, a linear stability analysis of the uniform state
reveals the mechanism leading to the pattern formation. The analysis of
non-linear effects is often realised numerically.
Weak Gaussian noise can select a particular pattern among many equivalent ones.

The theory of pattern formation was reviewed in~\cite{Cross-Hohenberg}. 
More recent developments in this field will be covered in the chapter 
by A. Nepomnyashchy.

\section{Beyond physics}

When modeling dynamical systems in fields of science other than
physics there is, basically, no constraint on the dynamic rules that can be used. 

The voter model is a particular reaction model, used to describe the spatial spreading
of opinions ~\cite{Castellano09} or  
populations~\cite{TilmanKareiva97} in terms of coarsening or segregation. A $q$-valued opinion variable is initially assigned, with some rule, 
to each site on a  lattice or a graph. 
The variables have no conviction and the microscopic dynamics is very simple: 
at each time step, a variable chosen at random adopts 
the opinion of a randomly-chosen neighbour. This parameter free  model approaches one of the  
$q$ absorbing states with complete consensus in a time that depends on the spatial dimensionality, the value of 
$q$ and the system size. The approach occurs via a coarsening process in $d\leq 2$ 
and because of a large random fluctuation in $d>2$. 
The coarsening process in $d=2$ is very different from the curvature driven one, demonstrating that symmetry alone does
not specify the asymptotic dynamics. A large bubble 
does not shrink but slowly disintegrates as its boundary roughens diffusively while the radius of the droplet remains 
statistically constant. These facts are associated to the absence of surface tension~\cite{Dornic-etal01}. 
The evolution of a random initial 
condition shows the growth of ordered spatial regions leading to growing length $R(t) \simeq t^{1/2}$ and the coarsening 
process is driven by interfacial noise.
In generalized versions of this models, with a temperature, a dynamic phase transition is defined as the critical
line between a low-temperature region where clusters
(or domains) grow indefinitely, and a high-temperature region where one only observes
fluctuations at a finite scale~\cite{DrouffeGodreche99}. 

\section{Quantum fluctuations}

A quantum system can undergo a quantum phase transition as a function of a parameter (say, 
external field or pressure) at strictly zero temperature. This 
critical point can continue into a critical line in the temperature - parameter of choice phase diagram. 

The relaxation dynamics of systems quenched to a quantum critical point 
were analysed in~\cite{Patane09,Znidaric-etal10} where various spin chains were considered,
and in~\cite{Gagel-etal14} where a vector field theory was studied
with renormalisation and large $N$ methods. 

An abrupt quench from the disordered into the ordered phase following some path in 
the, say, two-dimensional parameter space will set the system into a non-equilibrium initial 
condition and the equilibrium process can then occur via quantum coarsening.
There is no reason to believe that the dynamic scaling hypothesis will not  hold under quantum fluctuations. The reasonable 
assumption whereby fluctuations within domains should be determined by quantum and thermal 
fluctuations while the dynamics of interfaces should be close to the ones of their classical counterparts
has been demonstrated in some solvable mean-field models~\cite{Culo98,Aron-etal09}.
A scenario for the time-dependence of a growing length with several crossovers,  in the coarsening dynamics of 
$3d$ itinerant ferromagnets was put forward in~\cite{Belitz}.

Spin textures, ferromagnetic domains and vortices have been observed experimentally {\it in situ} 
in quenched Bose-Einstein condensates of atoms with non-zero internal angular momentum~\cite{Sadler-etal06}. The temporal and spatial 
evolution of these structures was also reported in this reference. The dynamics of this system is usually described with a 
Gross-Pitaevskii equation and numerical simulations confirm the domain growth observed experimentally.

\section{Wrup-up}

I have presented a short description of various systems evolving through phase ordering 
dynamics. In the rest of this volume some of these problems will be discussed in more 
detail.

\vspace{0.75cm}

\noindent
{\bf Acknowledgements.}
L. F. C. is a member of Institut Universitaire de France.

\vspace{1cm}

\bibliographystyle{phaip}
\bibliography{coarsening}

\end{document}